# SPECIAL SECTION ON STATISTICS IN THE ATMOSPHERIC SCIENCES

By Montserrat Fuentes, Peter Guttorp and Michael L. Stein

*North Carolina State University, University of Washington and University of Chicago*

With the possible exception of gambling, meteorology, particularly precipitation forecasting, may be the area with which the general public is most familiar with probabilistic assessments of uncertainty. Despite the heavy use of stochastic models and statistical methods in weather forecasting and other areas of the atmospheric sciences, papers in these areas have traditionally been somewhat uncommon in statistics journals. We see signs of this changing in recent years and we have sought to highlight some present research directions at the interface of statistics and the atmospheric sciences in this special section.

Two of the papers in this section relate to statistical approaches to precipitation modeling. The stochastic modeling of precipitation goes at least back to the introduction of Markov chains by Quetelet (1852) to describe dependent events of daily rainfall. In modern precipitation modeling there are three major strands: extensions of the Markov chain structure to hidden Markov models, a point process approach and spatial models based on Gaussian processes. Among these, the point process approach is most closely related to the physical structure of cyclonic storms.

The structure of cyclonic storms was studied in detail by Hobbs and Locatelli (1978). Each frontal system contains a sequence of rain bands, each containing rain cells of higher local precipitation intensity. In northern mid-latitudes the frontal systems in winter arrive at about a three day time scale. The systems are of a synoptic spatial scale of order of magnitude $10^3$ km, while the ensuing precipitation occurs on a mesoscale, $10^2$ km or less.

Le Cam (1961) assumed a directing measure $M$ that generates the random rate of another random measure $N$. The actual rainfall is then taken as a smoothing of the measure $N$. The point process approach was applied (in less generality than that of Le Cam) to precipitation data by Rodriguez-Iturbe, Cox, Foufoula-Georgiou and others in the 1980s. A more general approach was developed by Phelan (1996), who considered a stochastic flow to represent the atmosphere, within which rain cells were born and died according to birth and death processes. While this approach has the potential to describe precipitation on a synoptic scale, the statistical tools (as well as the appropriate data) were not available at the time. Generally, the lack of precise estimation tools for most cluster point processes has been hampering (and may have choked) developments in this area.

The Markov chain approach has been extended to spatial networks of stations by Zucchini, Hughes, Bellone and others. Here, a hidden Markov approach has been found an improved fit, and the hidden states can be thought of as precipitation regimes. Atmospheric covariates are allowed to affect transition probabilities, but there is no true spatial model that extends the network to a full spatial area. While the precipitation regimes often have a reasonable meteorological interpretation, the model lacks basis in hydrometeorology.

For fully spatial models one approach has been to use a Gaussian field for some transformation of rain measurements, while the negative part of the distribution gets truncated and corresponds to zero rain. Again, this model lacks physical basis. The paper by Berrocal, Raftery and Gneiting (2008) in this issue generalizes this approach by using an independent Gaussian process, the sign of which indicates precipitation occurrence, while the paper by Fuentes, Reich and Lee (2008) uses a gridded latent process.

It may be time to combine aspects of these different approaches into physically realistic descriptions of precipitation that can be used for downscaling climate models, for hydrologic modeling and for forecasting. Doing this in a way that allows for effective statistical inference and useful model diagnostics is likely to remain a challenge for the foreseeable future.

Models for measurements, both statistical and physical, are a critical issue in many areas of the atmospheric sciences. In particular, the measurement of precipitation is not an easy task. There are lots of devices, from tipping gauges to distrometers. No particular instrument gives an unbiased estimate of rain rate. Cardoso and Guttorp (2008) outline a hierarchical approach to combining the measurements from different devices to estimate the true underlying rate. The approach is based on the (unobserved) drop size distribution. In order to extend it to networks or spatial regions, one needs to model the spatial dependence of these distributions. The Fuentes, Reich and Lee (2008) paper avoids the physical modeling of the relation between radar reflectivity and gauge readings, replacing it with a statistical model. Remotely sensed observations, of which radar measurements are one example, provide for the opportunity of excellent spatial coverage, but may often exhibit spatial correlation and/or spatially dependent biases in their measurement errors, greatly complicating efforts to distinguish signal from noise.

Another challenge in making effective use of data in the atmospheric sciences is dealing with disparate data and model sources of information in a single analysis. Hierarchical models have proven effective in this regard [Bannerjee, Carlin and Gelfand (2004)] and are used in Shaddick et al. (2008) to estimate the effect of air pollution on human health. This problem is hugely challenging due to the small effect sizes at present pollution levels in the developed world and the difficulty of determining exposures at the individual level. The US Environmental Protection Agency now regularly estimates the health benefits of regulations affecting air pollution [National Research Council (2002)], so that determinations that there is an effect are not sufficient. Thus, it may not be adequate to ignore the distinction between pollution levels at monitoring sites and the actual exposures of individuals when estimating the effect of air pollution on human health.

Meteorological datasets and the output of meteorological models are often enormous, leading to a natural desire for dimension reduction methods. Thus, for example, principal components, generally called empirical orthogonal functions, or EOFs, in the atmospheric sciences, are commonly used to reduce descriptions of the state of the atmosphere to a modest number of dimensions [Wilks (2005)]. Similarly, a variety of clustering methods are often used to help atmospheric scientists summarize their masses of data and model output. Sang et al. (2008) consider self organizing maps, or SOMs, a clustering technique largely unknown to statisticians, as part of a space–time statistical model for atmospheric states over Southern Africa. Despite the usefulness of these various dimension reduction methods, from a statistical perspective, they are generally lacking in motivation. Specifically, other than trying a method on many examples, can one say when a particular method might be expected to work well? The fact that many such methods (such as SOMs and EOFs) make no explicit use of the spatial locations and times of the observations suggests that it might be possible to make sharper inferences about any underlying low-dimensional structures that may exist in such processes by taking account of spatial–temporal dependencies.

An essential ingredient to effective weather forecasting based on numerical models is the appropriate integration of observations into the initial conditions for the model, or data assimilation [Kalnay (2003)]. Objective analysis, which statisticians might call optimal linear interpolation or simple kriging, was an early approach to this problem [Gandin (1963), Thiébaux and Pedder (1987)]. Today, approaches based on various versions of the Kalman filter are popular. Statisticians who have passed through the National Center for Atmospheric Research (NCAR) as part of the Geophysical Statistics Project (GSP), notably including its first two heads, Mark Berliner and Doug Nychka, and their many postdocs, have played a major role in drawing statisticians into this line of work, which affords a wealth of statistical, computational and scientific challenges. The GSP is represented here by Malmberg et al. (2008), in which they explore the feasibility of combining observations and a numerical model for interpolating carbon monoxide fields on a large spatial scale. Thus, we see data assimilation moving from its traditional areas of application in meteorology and oceanography into atmospheric chemistry. Because there are more substantial scientific uncertainties (as opposed to uncertainties in initial and boundary conditions) in atmospheric chemistry than in atmospheric physics, taking proper account of these scientific uncertainties may be critical in obtaining realistic predictions and especially uncertainty estimates for these predictions.

Another widespread statistical issue in atmospheric science is inference for extremes, which is of critical importance in meteorology, hydrology and air quality monitoring. Spatial variability in extremes presents particular challenges, especially when, as is commonly the case, one wishes to make inferences about extremes at locations at which no monitoring data is available. In a paper that appeared in this journal, Buishand, de Haan and Zhou (2008) developed some theory for spatial extremes with an application to rainfall in Holland, although they needed to make quite strong assumptions on the nature of the spatial dependence to obtain results. There is considerable interest in how future changes in climate might impact extreme weather events. Making inferences about such quantities that include realistic assessments of uncertainty may seem too much to expect, but without at least some kind of answer to this question, how should one go about designing bridges or levees that are meant to last many decades?

The list of issues raised here only begins to touch on the many statistical problems in the atmospheric sciences that deserve further attention. Some others include statistical models for processes on a global scale [Jun and Stein (2008)], statistical models that capture dynamics, multivariate models for spatial–temporal processes and statistical models that include the vertical spatial dimension. Climate change, one of the great modern challenges to humanity, provides a myriad of statistical challenges, some of which are summarized in the American Statistical Association's recent Statement on Climate Change (2008). Drignei, Forest and Nychka (2008) describe an approach to estimating climate sensitivity (the change in global temperature due to a doubling of $CO_2$ concentration) that combines numerical climate model output with nonlinear regression and statistical methods for analyzing computer experiments. As in many areas, a major role for statisticians in climate change research is in appropriately quantifying uncertainties. Since climate change projections are necessarily statements about the climate under higher concentrations for greenhouse gases than have existed in historical times, such projections are extrapolations, which generally make statisticians apprehensive. Nevertheless, these kinds of daunting questions will be addressed with or without (but hopefully with) the statistical community's leadership. We look forward to publishing papers addressing this and other critical issues in the coming years.

M. Fuentes  
Department of Statistics  
North Carolina State University  
Raleigh, North Carolina 27695  
USA  
E-mail: fuentes@stat.ncsu.edu

P. Guttorp  
Department of Statistics  
University of Washington  
Seattle, Washington 98195  
USA  
E-mail: peter@stat.washington.edu

M. L. Stein  
University of Chicago  
Chicago, Illinois 60637  
USA  
E-mail: aoas@galton.uchicago.edu